# Generation of Chest CT pulmonary Nodule Images by Latent Diffusion Models using the LIDC-IDRI Dataset

Kaito Urata[1], Maiko Nagao[1], Atsushi Teramoto[1],

Kazuyoshi Imaizumi[2], Masashi Kondo[2], Hiroshi Fujita[3]

1 Graduate School of Science and Engineering, Meijo University

2 Fujita Health University, 3 Gifu University

## Abstract

**[Purpose]** Lung cancer is the leading cause of cancer-related deaths. Recently, computer-aided diagnosis (CAD) systems have been developed to support diagnosis, but their performance depends heavily on the quality and quantity of training data. However, in clinical practice, it is difficult to collect the large amount of CT images for specific cases, such as small cell carcinoma with low epidemiological incidence or benign tumors that are difficult to distinguish from malignant ones. This leads to the challenge of data imbalance. In this study, to address this issue, we proposed a method to automatically generate chest CT nodule images that capture target features using latent diffusion models (LDM) and verified its effectiveness.

**[Methods]** Using the LIDC-IDRI dataset, we created pairs of nodule images and finding-based text prompts based on physician evaluations. For the image generation models, we used Stable Diffusion version 1.5 (SDv1) and 2.0 (SDv2), which are types of LDM. Each model was fine-tuned using the created dataset. During the generation process, we adjusted the guidance scale (GS), which indicates the fidelity to the input text. The generated images were evaluated using both quantitative and subjective methods. Quantitative evaluation used metrics to measure image quality, diversity, and consistency with the text. Subjective evaluation consisted of visual assessments conducted by three experienced radiological technologists.

**[Results]** Both quantitative and subjective evaluations showed that SDv2 (GS = 5) achieved the best performance in terms of image quality, diversity, and text consistency. In the subjective evaluation, no statistically significant differences were observed between the generated images and real images, confirming that the quality was equivalent to real clinical images.[

**[Conclusion]** We proposed a method for generating chest CT nodule images based on input text using LDM. Evaluation results demonstrated that the proposed method could generate high-quality images that successfully capture specific medical features.

## 1. Introduction

Lung cancer is the leading cause of cancer death worldwide [1]. As the disease progresses, prognosis rapidly worsens, making early detection and treatment essential. The national lung screening trial (NLST) demonstrated that lung cancer screening using CT images improves patient outcomes [2]. While CT images provide high sensitivity for detecting pulmonary nodules and observing small lesions, distinguishing benign from malignant nodules based on these images alone remains difficult. According to the results of NLST, 96.4% of cases requiring further investigation in the low-dose CT group were benign [2]. Such false positives can lead to unnecessary tests, biopsies, and invasive procedures, such as surgical resection [3]. Therefore, accurately differentiating between benign and malignant nodules is crucial for reducing patient burden and determining treatment strategies. Recently, computer-aided diagnosis (CAD) systems using deep neural networks (DNNs) have been developed to support physician's diagnosis [4]. Their performance, however, depends heavily on the quantity and quality of training data. Collecting large datasets for specific cases—such as rare small cell carcinoma or benign tumors—is challenging, often leading to biased data. To address this, Text-to-Image (T2I) technology has emerged, enabling the generation of diverse images that capture desired features. In this study, we used latent diffusion models (LDM) [5] to generate images and evaluated their effectiveness, including image quality.

## 2. Materials and Methods
### 2.1. Overview

Fig.1 shows an overview of the proposed method. In this study, we created image and text data from the lung image database consortium and image database resource initiative (LIDC-IDRI) dataset [8]. Using these as training data, we fine-tuned Stable Diffusion (SD), a type of latent diffusion model (LDM). Finally, we conducted both quantitative and subjective evaluations of the generated images.

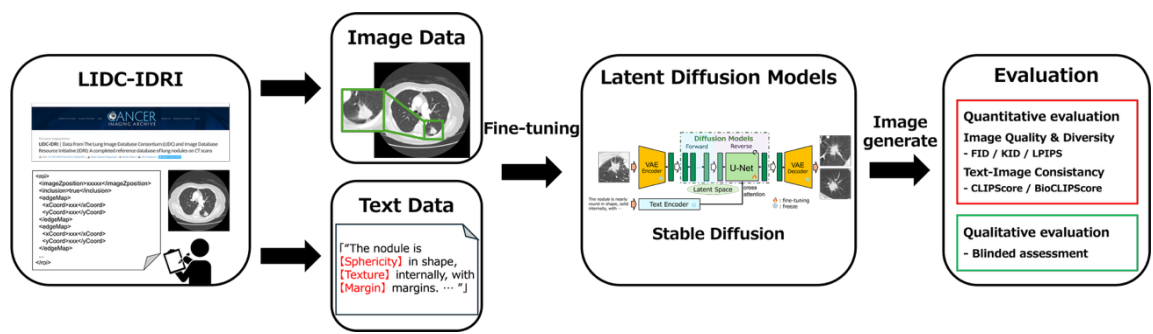

Fig.1 Overview of proposed method

## 2.2. Dataset Construction

We used the LIDC-IDRI chest CT database [8] for our training dataset. This database contains xml files with information on the location and morphological features of lung nodules. In this study, we selected the center slice along the Z-axis for each nodule. From these slices, we made the region of interest (ROI) image from original CT images, where each side of the image was twice the length of the nodule's maximum diameter. The corresponding text data was generated from the morphological feature scores provided by physicians in the xml files. We extracted scores for sphericity, margin, texture, spiculation, and the presence of calcification. These numerical scores were then converted into natural language descriptions and combined to create the final text prompts. Fig.2 shows examples of the training dataset. We split the dataset into training, validation, and test sets in a ratio of 7:2:1 (1,453, 416, and 208 images, respectively), ensuring that malignancy levels were evenly distributed across the sets. To prevent overfitting, we applied data augmentation to the training set, including rotation (90°, 180°, and 270°) and horizontal flipping. These augmented image-text pairs were then used for training.

| Images | prompts |
|---|---|
| Malignancy : 1 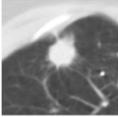 | The nodule is round in shape, solid internally, with well-defined margins. |
| Malignancy : 3 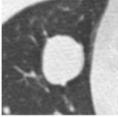 | The nodule is nearly round in shape, solid internally, with mostly well-defined margins. |
| Malignancy : 5 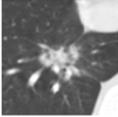 | The nodule is oval in shape, solid internally, with relatively well-defined margins. marked spiculation is seen. |

Fig.2 Examples of image-text pairs used for training

## 2.3. Image Generation

We adopted SD as our image generation model, which enables the generation of images from text prompts. Its architecture is illustrated in Fig. 3. SD is a type of LDM based on the diffusion models proposed by Sohl-Dickstein et al. and Ho et al. [9, 10]. In this study, we utilized two versions of the model: version 1.5 [6] and 2.0 [7] (SDv1 and SDv2).

SD utilizes a variational autoencoder (VAE) to compress images into a latent space, allowing for the sampling of high-resolution images while minimizing computational costs. To incorporate text-based conditions, a CLIP text encoder is employed. The extracted text features are fed into the cross-attention layers within the U-Net, enabling image generation that aligns with the provided text content.

In this study, to adapt these models to the lung nodule CT image domain, we performed full fine-tuning on the U-Net, which is the core image-generating component of SD. The implementation was carried out using the PyTorch deep learning framework and the Hugging Face Diffusers library. During training, all images were resized to a matrix size of $512 \times 512$ pixels. We used the AdamW optimizer with a learning rate set to $1.0 \times 10^{-6}$. Based on the convergence of the KID score on the validation data, training was terminated at 80 epochs for SDv1 and 90 epochs for SDv2.

During the image generation phase, we adjusted the guidance scale (GS), a parameter that controls the fidelity of the generated image to the input text. A lower GS value increases image diversity, whereas a higher value strengthens the alignment with the text prompt. However, excessively high values can lead to image degradation, such as the appearance of artifacts. In this study, we generated images using seven levels of guidance scale: GS = 5, 10, 20, 30, 40, 50, and 60.

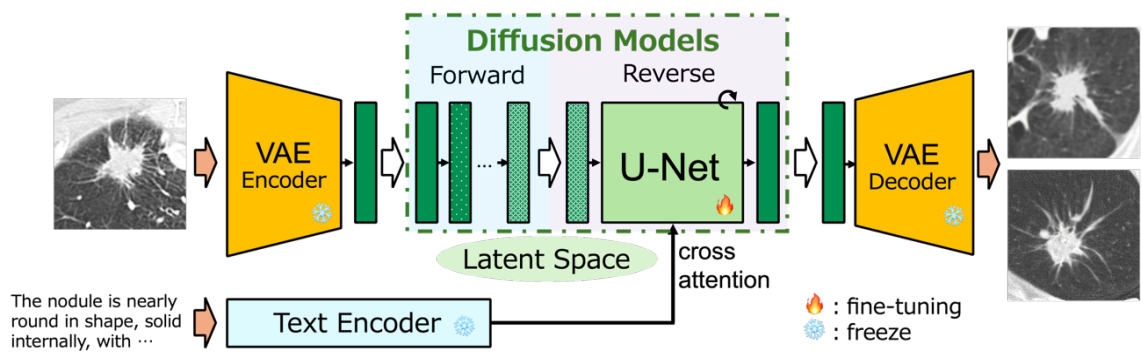

Fig.3 Stable Diffusion architecture

## 2.4. Evaluation strategy

Five quantitative metrics and subjective evaluation by three radiological technologists were used for the assessment.

First, in the quantitative evaluation, Fréchet inception distance [11] (FID) and kernel inception distance [12] (KID) were used to evaluate overall image quality and diversity, and learned perceptual image patch similarity [13] (LPIPS) was used to measure perceptual similarity. CLIPScore [14] was adopted to evaluate the semantic consistency with the text. Furthermore, considering that CLIPScore has limitations in understanding medical terminology as it is a model trained on general images, BiomedCLIP [15] trained on PMC-15M was introduced to evaluate the reproducibility of medical features of the nodules. This metric is referred to as BioCLIPScore. In the calculation of each CLIPScore, the values obtained by multiplying the cosine similarity by a scaling factor (w = 2.5) were used.

Meanwhile, in the subjective evaluation, visual assessment by three experienced radiological technologists was conducted to verify the clinical utility of the generated images. For the evaluation images, we prepared 20 cases each of real images and images of SDv1 and SDv2 generated from the same findings as the real images (60 images in total). The evaluation items consisted of seven categories: Realism, Malignancy, Sphericity, Texture, Margin, Spiculation, and Lobulation. For the evaluation results, the mean and standard deviation of each item were calculated, and the Mann-Whitney U test was used to determine significant differences.

3. **Results and Discussion**

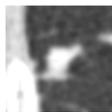

Fig.3 Examples of pulmonary nodule images generated by SDv1, SDv2 at GS = 5

Examples of lung nodule images generated by SDv1 and SDv2 are shown in Fig.3. The generated images capture the characteristics of both benign and malignant nodules. Benign nodules are often characterized by a round shape, and the model successfully generated malignant images with spiculated features. As shown in the quantitative evaluation results in Table 1, SDv2 (GS = 5) achieved the best performance. The image quality metrics (FID, KID, and LPIPS) reached their lowest (best) values, and the generated images exceeded the scores obtained when each CLIPScore was applied to real images (0.617 and 0.840). These results confirm that high-quality images reflecting the text prompts were successfully generated.

Table 1 Evaluation of quantitative metrics for generated pulmonary nodule images

| **Metrics** | **Models** | **GS5** | **GS10** | **GS20** | **GS30** | **GS40** | **GS50** | **GS60** |
|---|---|---|---|---|---|---|---|---|
| FID(↓) | SDv1 | 114.04 | 115.0 | 132.2 | 155.6 | 184.0 | 214.6 | 240.5 |
|  | SDv2 | **96.34** | 103.1 | 133.0 | 202.9 | 268.9 | 308.7 | 326.4 |
| KID(↓) | SDv1 | 0.063 | 0.059 | 0.070 | 0.089 | 0.119 | 0.154 | 0.189 |
|  | SDv2 | **0.038** | 0.039 | 0.056 | 0.129 | 0.208 | 0.257 | 0.280 |
| LPIPS(↓) | SDv1 | 0.449 | 0.452 | 0.461 | 0.470 | 0.483 | 0.496 | 0.508 |
|  | SDv2 | **0.441** | 0.449 | 0.466 | 0.480 | 0.500 | 0.525 | 0.551 |
| CLIPScore(↑) (w = 2.5) | SDv1 | 0.657 | 0.644 | 0.653 | 0.656 | 0.650 | 0.644 | 0.642 |
|  | SDv2 | **0.663** | 0.642 | 0.587 | 0.569 | 0.568 | 0.565 | 0.562 |
| BioCLIPScore(↑) (w = 2.5) | SDv1 | 0.835 | 0.848 | 0.801 | 0.762 | 0.736 | 0.723 | 0.718 |
|  | SDv2 | **0.870** | 0.854 | 0.778 | 0.726 | 0.687 | 0.643 | 0.618 |

Furthermore, in the subjective evaluation results shown in Table 2, the scores for each item in SDv2 were close to those of the real images. A comparison using the Mann-Whitney U test revealed no statistically significant differences between SDv2 and the real images across all seven evaluation categories.

Table 2 Subjective evaluation of real and generated images by radiological technologists

| Metrics | Real | SDv1 | SDv2 |
|---|---|---|---|
| Realism | 3.48 ± 1.01 | 3.05 ± 1.19 | 3.47 ± 1.19 |
| Malignancy | 3.48 ± 1.15 | 2.88 ± 1.07 | 3.48 ± 1.20 |
| Sphericity | 3.47 ± 1.15 | 3.70 ± 1.08 | 3.47 ± 1.16 |
| Texture | 3.47 ± 1.06 | 3.70 ± 1.00 | 3.68 ± 0.99 |
| Margin | 3.57 ± 1.13 | 3.48 ± 1.08 | 3.72 ± 1.07 |
| Spiculation | 3.27 ± 1.17 | 3.30 ± 1.11 | 3.23 ± 1.33 |
| Lobulation | 3.45 ± 0.92 | 3.60 ± 0.94 | 3.60 ± 0.94 |

Based on the quantitative and subjective evaluation results, it was confirmed that this method can generate high-quality images that accurately reflect the text content and are clinically realistic. In the future, we plan to utilize the generated images as training data and verify their effectiveness in improving the accuracy of image classification and lesion detection models. Furthermore, we aim to improve the performance of CAD systems by expanding the method to generate whole CT images and addressing a wider variety of cases.

## 4. Conclusions

In this study, we proposed the automatic generation of chest CT nodule images using LDM with the LIDC-IDRI dataset. Our verification showed that we could generate a large quantity of diverse images that successfully captured the target features. These results suggest that the proposed method is not limited to lung nodules and has the potential to be applied to various other diseases as well.

**Acknowledgements**

This research was supported in part by a Grant-in-Aid for Scientific Research (No. 23K07117)

from the Ministry of Education, Culture, Sports, Science and Technology, Japan.

**Conflict of Interest**

The authors declare that they have no conflict of interest.

**Ethical approval**

This study was performed using the LIDC-IDRI, a publicly available database. The data were anonymized before being made available, and ethical approval for the original data collection was obtained by the source institutions.

**Informed consent**

For this type of study, formal consent is not required as the data were obtained from an anonymized public database.

## References

[1]  American Cancer Society (2025) Cancer facts and figures 2025. American Cancer Society, Atlanta

[2]  National Lung Screening Trial Research Team (2011) Reduced lung-cancer mortality with low-dose computed tomographic screening. N Engl J Med 365(5):395–409

[3] Asano F, Aoe M, Ohsaki Y, Okada Y, Sato S, Ohmichi M, Saka H, Itoh K, Kaneko K, Sasada S (2012) Deaths and complications associated with respiratory endoscopy: a survey by the Japan Society for Respiratory Endoscopy in 2010. Respirology 17(3):478–485

[4] Chan HP, Hadjiiski LM, Samala RK (2020) Computer-aided diagnosis in the era of deep learning. Med Phys 47(5):e218–e227

[5] Rombach R, Blattmann A, Lorenz D, Esser P, Ommer B (2022) High-resolution image synthesis with latent diffusion models. In: Proc IEEE/CVF Conf Comput Vis Pattern Recognit, pp 10684–10695

[6] RunwayML (2022) Stable Diffusion v1.5. [https://huggingface.co/runwayml/stable-diffusion-v1-5](https://huggingface.co/runwayml/stable-diffusion-v1-5). Accessed 3 Jan 2026

[7] Stability AI (2022) Stable Diffusion v2.0. [https://huggingface.co/stabilityai/stable-diffusion-2](https://huggingface.co/stabilityai/stable-diffusion-2). Accessed 27 Dec 2025

[8] Armato III SG, McLennan G, Bidaut L, McNitt-Gray MF, Meyer CR, Reeves AP, Zhao B, Aberle DR, Henschke CI, Hoffman EA, Kazerooni EA, MacMahon H, Van Beek EJR, Croft BY, Clarke LP (2011) The Lung Image Database Consortium (LIDC) and Image Database Resource Initiative (IDRI): a completed reference database of chest CT scans on cancer screening. Med Phys 38(2):915–931

[9] Sohl-Dickstein J, Weiss EA, Maheswaranathan N, Ganguli S (2015) Deep unsupervised learning using nonequilibrium thermodynamics. In: Proc 32nd Int Conf Mach Learn, pp 2256–2265

[10] Ho J, Jain A, Abbeel P (2020) Denoising diffusion probabilistic models. In: Adv Neural Inf Process Syst, pp 6840–6851

[11] Heusel M, Ramsauer H, Unterthiner T, Nessler B, Hochreiter S (2017) GANs trained by a two time-scale update rule converge to a local Nash equilibrium. In: Adv Neural Inf Process Syst, pp 6626–6637

[12] Bińkowski M, Sutherland DJ, Arbel M, Gretton A (2018) Demystifying MMD GANs. In: Proc 6th Int Conf Learn Represent

[13] Zhang R, Isola P, Efros AA, Shechtman E, Wang O (2018) The unreasonable effectiveness of deep features as a perceptual metric. In: Proc IEEE Conf Comput Vis Pattern Recognit, pp 586–595

[14] Hessel J, Holtzman A, Forbes M, Le Bras R, Choi Y (2021) CLIPScore: a reference-free evaluation metric for image captioning. In: Proc Conf Empir Methods Nat Lang Process, pp 7514–7528

[15] Zhang S, Xu Y, Usuyama N, Bagga J, Tinn R, Preston S, Howard R, Naumann F, Gao J, Poon H (2023) BiomedCLIP: a multimodal biomedical foundation model pretrained from fifteen

million scientific image-text pairs. arXiv preprint arXiv:2303.00915